\RequirePackage{iftex}
\ifPDFTeX
  \pdfoutput=1
\fi

\documentclass{article}
\usepackage{arxiv}

\usepackage[utf8]{inputenc}
\usepackage[T1]{fontenc}
\usepackage[strings]{underscore}
\usepackage{hyperref}
\usepackage{booktabs}
\usepackage{amsmath}
\usepackage{amssymb}
\usepackage{graphicx}
\usepackage{microtype}
\usepackage{natbib}
\usepackage{minted}
\setminted{breaklines=true}

\usepackage{xcolor}
\usepackage{tcolorbox}
\tcbuselibrary{skins,breakable}

\newtcolorbox{authorversionbox}{
  colback=gray!8,
  colframe=gray!60,
  boxrule=0.4pt,
  arc=1mm,
  left=6pt, right=6pt, top=4pt, bottom=4pt,
  breakable,
  fontupper=\small
}

\renewcommand{\headeright}{Preprint}
\renewcommand{\undertitle}{Author's version --- accepted at SWODCH 2026, co-located with ISWC 2026}
\renewcommand{\shorttitle}{Preserving Contextual Information via Multidimensional Knowledge Graphs}

\title{Preserving contextual information in cultural heritage metadata through multidimensional knowledge graphs}

\author{
  Lyndon Nixon$^{\dagger,*}$ \\
  Storypact GmbH, Vienna, Austria \\
  and Modul University Vienna, Austria \\
  \And
  Valentina Presutti$^{\dagger}$ \\
  University of Bologna \\
  \And
  Celian Ringwald$^{\dagger}$ \\
  University of Bologna \\
  \And
  Andrea Schimmenti$^{\dagger}$ \\
  University of Bologna \\
}

\date{}

\begin{document}
\maketitle

{\renewcommand{\thefootnote}{}\footnotetext{$^{*}$Corresponding author. \quad $^{\dagger}$These authors contributed equally.}}

\begin{authorversionbox}
\noindent\textbf{This is the authors' version of the work.} It is posted here for personal use, not for redistribution. The definitive version was published in the \emph{Proceedings of SWODCH 2026: the 6th International Workshop on Semantic Web and Ontology Design for Cultural Heritage}, co-located with ISWC 2026, Bari, Italy, 25--26 October 2026, published by CEUR Workshop Proceedings (CEUR-WS.org), Vol.~XXXX, and licensed under CC BY 4.0. 
\end{authorversionbox}
\bigskip

\begin{abstract}
Using Knowledge Graphs (KGs) to describe Cultural Heritage Objects (CHOs) supports semantic richness and interoperability. However, standard KGs fail to capture the context-dependent validity of statements. This limitation is critical for cultural heritage metadata, which must often accommodate evolving or conflicting viewpoints, such as colonial versus post-colonial perspectives or shifting scientific consensus. While current knowledge representation methods address basic contextualization via provenance, qualifiers or reification, they lack a unified framework to simultaneously model and query data across multiple social, cultural, and political dimensions. To bridge this gap, we introduce the conceptual foundations of Multi-dimensional Knowledge Graphs (MKGs) and discuss how they preserve complex, multi-layered contexts in CHO metadata.
\end{abstract}


\section{Introduction}\label{sec:intro}

Knowledge Graphs (KGs) are an important mechanism for representing and interlinking Cultural Heritage (CH) knowledge \cite{10.1145/3607542.3617354}. Through the reuse of shared entities, vocabularies, and ontologies, they enable semantic interoperability across collections, institutions, and domains. Widely adopted CH models such as CIDOC CRM ~\cite{doerr_cidoc_2003} encourage the representation of Cultural Heritage Objects (CHOs) as interconnected networks of
entities and relationships, allowing object metadata to be linked to broader historical, geographical, and social contexts. The EU funded INFINITY project \cite{infinity_consortium_infinity_2025} has as one of its goals the promotion of CHO information representation via ontologies and knowledge graphs, preserving contextual interpretations and shared via the ECCCH (European Collaborative Cloud for Cultural Heritage). 

However RDF's model-theoretic semantics does not specify under what conditions an asserted triple should be considered true, leaving a statement's truth --- objective fact, situated belief, or something contingent on a particular standpoint --- entirely outside the formal model. This absence motivates attempts to introduce contextual reasoning into RDF knowledge bases \cite{bouquet_introducing_2005}. Existing work provides mechanisms for
provenance~\cite{lebo_prov-o_2013}, uncertainty~\cite{udrea_annotated_2010}, and
annotations~\cite{hartig_foundations_2017}, but
they offer limited support for representing situations in which the validity or interpretation of a statement depends on a particular social, cultural, political, temporal, or spatial perspective.

CHOs often embody multiple and sometimes conflicting
interpretations that coexist across communities and historical
periods \cite{amsdottorato12142}. A colonial-era map may describe a territory using names and
political boundaries that differ from contemporary understandings.
Ethnographic collections may employ classifications that are no longer
accepted by the communities represented \cite{turner_cataloguing_2020}. Historical
scientific instruments may encode cosmological models that have since been
superseded. In such cases, preserving CH requires preserving not only the
object and its metadata, but also the contextual frameworks through which
that metadata was originally understood.

Consider the example of the city commonly known as City of Derry in
Northern Ireland. The name used to refer to the city differs
depending on the socio-cultural and political identity of the speaker
(British Unionists still preferring Londonderry) --- a clear illustration of the
politics of toponymic inscription documented in the critical place-name
studies literature\cite{rose-redwood_geographies_2010}.

Similarly, statements concerning the status of territories such as Northern
Cyprus or Crimea may vary according to geopolitical perspectives. Traditional
knowledge graph models typically resolve such situations either by
identifying a preferred statement, by recording multiple statements
without explicit contextual semantics, or by relying on provenance metadata that identifies who asserted a claim without formally
capturing the conditions under which that claim is considered
valid. Consequently, applications built upon these graphs are unable to
systematically query or navigate knowledge from different contextual
viewpoints.

To address this challenge, we propose the concept of  Multidimensional
Knowledge Graph (MKG). MKGs leverage RDF 1.2 by allowing
individual statements to be contextualised through explicit dimensions of interpretation, building on prior annotation-based
contextualisation
frameworks \cite{udrea_annotated_2010,gimenez-garcia_ndfluents_2016-1,gimenez-garcia_ndfluents_2017}.

Examples of dimensions include temporal periods, geographical regions,
political perspectives, cultural communities, linguistic groups, or any
other epistemic category relevant to a domain.

Rather than associating a statement with a single contextual value (e.g., provenance), a
MKG represents constraints over a dimension's value space, enabling
statements to be expressed as valid for specific groups, valid except for
particular groups, or valid within defined temporal or spatial extents.

Building upon RDF 1.2~\cite{w3c_rdf-star_working_group_rdf_2025}, formerly RDF-star~\cite{hartig_foundations_2017}, contextual information can be attached directly to individual statements while remaining compatible with existing Linked Data infrastructures by following the Expressing Without Asserting (EWA) principles\cite{daquino_expressing_2022}.

The proposed approach offers several potential benefits for CH knowledge
representation. First, it enables multiple coexisting statements about the same object or concept to be preserved without requiring a single
authoritative viewpoint. Second, it supports context-aware querying and
exploration, allowing users to navigate knowledge from the perspective of particular communities, cultures, or historical periods.
Third, it provides a mechanism for representing contested, evolving, or
pluralistic knowledge in a structured and machine-readable form. Such
capabilities are increasingly important as CH institutions seek to
incorporate diverse perspectives and address historical biases within their
collections\cite{turner_cataloguing_2020}.

Nevertheless, the introduction of multidimensionality raises
several challenges. Additional contextual structures increase modelling
complexity and may impact query performance and
scalability\cite{carroll_named_2005}. The definition of dimensions and their
value spaces introduces new ontology engineering requirements, particularly
when dimensions require hierarchical organisation and
reasoning\cite{bozzato_materialization_2013}. Furthermore, questions remain regarding
governance, interoperability, and the potential proliferation of competing
contextual statements/interpretations. Determining which dimensions should be
represented and how they should be maintained may become a socially and
politically contested process\cite{turner_cataloguing_2020,rose-redwood_geographies_2010}.

In this paper, we reflect on previous work on preserving contextual interpretations of statements in
knowledge graphs (Section.~\ref{sec:background}). We then present the conceptual foundations of
Multidimensional Knowledge Graphs (MKG) and discuss how they can support the
preservation of arbitrary contexts of statements in CHO metadata
(Section.~\ref{sec:mkg}). We illustrate the approach through examples drawn from cultural heritage
(Section.~\ref{sec:examples}), and we examine the opportunities and limitations of introducing
multidimensional contextualisation into KG representations 
(Section.~\ref{sec:discussion}). We conclude with
our plan to establish MKGs in the cultural heritage domain through the
INFINITY project
(Section.~\ref{sec:conclusion}).

\section{Background and Related Work}\label{sec:background}

Semantic web ontologies and knowledge graphs rely on formal logic (RDF Schema, OWL), whose model-theoretic semantics is silent regarding the conditions under which a statement holds. This becomes an obstacle when facts need qualification beyond simple binary predicates \cite{bouquet_introducing_2005}. The use of these technologies in domain-specific applications and in general-purpose resources, such as DBPedia and Wikidata, made this limitation evident: facts differ within the same interpretative context, such as in political conflicts or because of spatial, temporal or socio-cultural viewpoints.

Some CH infrastructures tackle this issue by addressing multi-perspective institutional representations: the Europeana Data Model introduces the notion of proxy\cite{isaac_europeana_2013,noauthor_europeana_nodate}, allowing contributing institutions to assert their own version of an object's metadata within a shared aggregation, rather than forcing convergence on a single authoritative record. This approach is a step towards MKG but it is insufficient as it operates on whole records rather than individual statements, and leaves out non-institutional dimension, such as time, geography, or political perspectives.

\subsection{Context and Statement-Level Metadata in RDF}\label{sec:reification}

RDF reification, introduced in the original RDF specifications\cite{hayes_rdf_2004,manola_rdf_2004}, is a way to represent RDF statements' contextualisation. It enables metadata to be attached to statements, such as provenance information, confidence values, or annotations. It has very limited practical adoption due to its verbosity and lack of a formally defined interpretation linking the reified statement to the original assertion~\cite{hartig_foundations_2014}.

The W3C PROV-O ontology\cite{lebo_prov-o_2013} focuses on modeling provenance: statements or datasets are associated with agents, activities, and sources, thereby recording who asserted a claim and how it was produced. Provenance is essential for establishing trust and traceability and is frequently used in cultural heritage and scientific data management\cite{groth_anatomy_2010}.

A recent survey scoped on cultural heritage metadata identifies Named Graphs, RDF-star, PROV-O, and related models as the most effective solutions for representing provenance and tracking changes in RDF. It  highlights trade-offs in standards compliance, scalability, and cross-domain applicability discussed throughout this section~\cite{10.1093/llc/fqaf076}.

We argue that statement annotation fail to provide a general theory of context or interpretation. Unless the modelling pattern explicitly distinguishes the information's validity, metadata attached to a statement is insufficient to determine how that statement contributes to the truth conditions of a dataset.

\subsection{Named Graphs and Dataset-Level Context}\label{sec:named-graphs}

Named graphs~\cite{carroll_named_2005} group RDF triples into distinct graphs within an RDF dataset. A named graph can be used as the target of metadata, allowing information such as provenance, authorship, temporal scope, trust, access rights, or dataset boundaries to be attached to a collection of triples rather than to each individual statement. They are used to represent provenance, trust, temporal snapshots (e.g.\ \cite{hofer_dbpedia-tkg_2025}) and are central to the nanopublication model~\cite{groth_anatomy_2010}. In CH and digital
hermeneutics they are relevant because interpretative claims (attribution, interpretation, or scholarly hypothesis) involve several interdependent statements. Projects such as MythLOD\cite{Pasqual_Tomasi_2022}
and mAuth\cite{Daquino_2017} use domain ontologies and named graphs to represent artwork interpretations and attributions. Named graph provide a technical solution for grouping statements but the semantics of what they represent --- assertions, quotes, hypotheses, etc. --- remains implicit or delegated to external means (ontologies, application profiles, etc.). Moreover, when contextualisation is required statement-level, named graphs may force the creation of many small fragments, increasing management complexity. This is especially problematic when a dataset must represent different kinds of contextual information simultaneously, such as provenance, temporal validity, interpretative stance, evidential basis, and assertion status.

\subsection{Qualifiers and Statement Annotation in Knowledge Graphs}\label{qualifiers}

Large-scale knowledge graph projects have developed their own mechanisms for statement-level contextualisation. Wikidata introduced
\emph{qualifiers} to attached contexts to individual
claims\cite{vrandecic_wikidata_2014}. Qualifiers are used to express temporal validity, spatial scope, measurement conditions, and other constraints on a statement. This approach demonstrated the practical value of contextualised statements at web scale. However, qualifiers are application-specific and are outside RDF semantics.
Comparative studies between qualifiers and RDF reification show that singleton properties\cite{nguyen_dont_2014} and n-ary relation
patterns\cite{noy_defining_2006} offer different trade-offs for capturing statement-level qualification\cite{hernandez_reifying_2015}.  

\subsection{RDF-star and RDF 1.2}\label{sec:rdfstar}

To address the practical limitations of reification, RDF-star introduces a concise syntax for expressing statements about statements through quoted triples\cite{udrea_annotated_2010}, RDF 1.2~\cite{w3c_rdf-star_working_group_rdf_2025} enables metadata to
be attached directly to individual assertions reducing the  modelling complexity and has been adopted by several graph databases and Semantic Web tools. RDF 1.2 provides a standardised foundation for annotating triples and expressing metadata such as provenance, confidence, temporal validity, and other contextual information. Nevertheless, RDF-star intentionally remains agnostic regarding the semantics of such annotations and does not prescribe how contextual validity should be represented,
interpreted, or queried across different dimensions.

RDF 1.2 also makes it possible to express a statement without asserting (EWA) it: a quoted triple term can be used as the target of metadata without entailing that the corresponding triple is asserted in the graph\cite{daquino_expressing_2022}. While this allows statements to be expressed without committing to a specific truth value, the contextual validity of the triple has to be modelled outside.

\subsection{Property Graphs and Graph Databases}\label{sec:propertygraphs}

Outside the Semantic Web ecosystem, property graph models, implemented by
systems such as Neo4j, TigerGraph and JanusGraph, allow 
properties to be attached directly to both nodes and
edges\cite{robinson_graph_2015,angles_property_2018}. This capability has made property graphs attractive for representing contextual information and graph analytics applications. Contextual attributes such as timestamps,
confidence values, or provenance metadata can be attached directly to
relationships without requiring additional graph structures.
While property graphs provide an intuitive model for edge-level
annotations, they lack formal semantics, ontology alignment mechanisms,
and Linked Data interoperability provided by RDF. Furthermore, contextual
attributes are typically treated as implementation-level properties rather
than components of a formally defined framework for reasoning about
contextual validity.

\subsection{Temporal, Spatial, and Contextual Knowledge Graphs}\label{sec:temporal-spatial}

A substantial body of research has explored the representation of
specific forms of contextual knowledge. Temporal knowledge graphs model
facts that evolve over time, often through temporal intervals associated
with entities or
relationships\cite{welty_reusable_2006,gutierrez_introducing_2007}
Such approaches support temporal reasoning and historical querying but
focus primarily on a single dimension of context. Similarly,
GeoSPARQL\cite{battle_enabling_2012} and related geospatial extensions
enable the representation and querying of geographically scoped knowledge.
These technologies provide rich spatial reasoning capabilities, including
topological relations and geometric containment. However, they address
spatial validity only and do not generalize to other forms of contextual
interpretation.

More recently, research on contextualized knowledge
graphs, knowledge graph embeddings with context, and multiview knowledge
representations has investigated the incorporation of provenance, trust,
uncertainty, perspectives, and situational factors into graph structures. Formal frameworks such as
Annotated RDF \cite{udrea_annotated_2010}, Contextualized Knowledge
Repositories\cite{bozzato_materialization_2013} and
Context-OWL\cite{bouquet_c-owl_2003}, as well as multi-dimensional extensions of the fluents pattern such as
NdFluents\cite{gimenez-garcia_ndfluents_2016-1,gimenez-garcia_ndfluents_2017}, have also been proposed.

These approaches acknowledge that knowledge may be dependent on context and have demonstrated the value of context-aware retrieval and reasoning.
Nevertheless, many remain specialized to particular domains, rely on
application-specific representations, or lack a unified framework capable
of expressing multiple contextual dimensions simultaneously.

\subsection{Perspective and Argumentation Models}\label{sec:perspective}

A further line of work addresses contextualization along a single
perspectival dimension rather than time or space. Linguistic resources
such as Connotation Frames\cite{rashkin_connotation_2016} annotate
frame-semantic roles with polarity, capturing how a statement casts its
participants in a positive or negative light depending on the speaker's
stance --- an early attempt to formalize ``point of view'' as a structured
annotation rather than free text. Frame-centric linked data hubs such as
Framester\cite{gangemi_framester_2016} provide the underlying frame/role
infrastructure that such perspective layers can be attached to. In the
fact-checking domain, schema.org's ClaimReview and the Argument
Interchange Format\cite{chesnevar_towards_2006} provide structured ways
of expressing that a statement's reliability or argumentative status
depends on who is asserting and reviewing it. Closer to the cultural heritage
domain, the MULTI ontology\cite{gillis-webber_towards_2023}
models the natural-language and cultural viewpoint from which a
(multilingual) domain ontology has been authored, annotating ontology
entities with the linguistic or cultural community whose perspective they
encode. In digital humanities, Masolo et
al.\cite{masolo_observational_2025} propose an ontology of
\emph{observations} that reifies scholars' interpretive claims about
literary texts, tracking each claim's source observer and modelling
\emph{assertion}, \emph{denial}, \emph{support}, and \emph{defeat}
relations so that conflicting interpretations --- e.g., diverging critical
readings of the same character --- can coexist and be compared without
commitment to a single valid reading. As with the temporal and
spatial extensions above, these approaches are effective within their
target dimension but remain domain-specific, with no general mechanism for
combining their dimensions with, e.g., a temporal or
political one.

\subsection{Remaining Gaps}\label{sec:gaps}

Taken together, existing approaches have significantly advanced the ability to
represent metadata/context about knowledge statements. Reification and RDF-star provide
mechanisms for statement-level annotation. Provenance models identify
sources and agents. Named graphs introduce graph - level contexts.
Qualifiers support practical contextualization in large knowledge graphs.
Temporal and spatial extensions enable reasoning over specific dimensions
of validity. A smaller number of approaches have sought generality rather
than a single dimension. NdFluents extends the 4D-Fluents pattern to an
arbitrary number of context dimensions, Annotated RDF generalizes triple
annotation to a lattice-based algebra over multiple annotation domains,
and Contextualized Knowledge Repositories provide a Description Logic
semantics for reasoning across explicitly nested contexts.

However, a common limitation remains: contextual information is
typically represented as isolated annotations whose semantics are defined
implicitly by individual applications rather than through a general model
of contextual validity, and none of these proposals were designed
against the statement-level reification mechanism that RDF itself has since
standardized.
Existing approaches provide mechanisms for attaching metadata to statements
but do not offer a unified framework for expressing that a statement is
valid only under specific social, cultural, political, temporal, spatial,
linguistic, or other epistemic conditions. Nor do they provide a general
mechanism for querying and reasoning across heterogeneous dimensions of
context.

The need to make a statement's context of validity explicit is therefore not itself new: it motivated some of the earliest proposed extensions to RDF semantics\cite{bouquet_introducing_2005}, well before the formal frameworks surveyed above. What has changed is the underlying infrastructure available to express that need. The contribution we pursue here is accordingly narrower than introducing context to RDF: it is a multidimensional representation of context expressible natively within, and validated by, the now-standardised RDF 1.2 reification mechanism, rather than requiring a bespoke semantic extension.
The emergence of RDF 1.2 provides an
opportunity to address these limitations. Rather than replacing existing Semantic Web technologies, the Multidimensional Knowledge
Graph (MKG) seeks to
complement them by providing a unified
and extensible framework for representing and querying contextual
interpretations while remaining fully compatible with RDF 1.2, Linked Data
principles, and existing RDF-based infrastructures.

\section{Multidimensional Knowledge Graph (MKG) Model}\label{sec:mkg}
\subsection{Epistemological Motivation}\label{sec:epistemology}

Knowledge Graphs are typically built atop RDF, whose semantics leaves the validity conditions of a statement unspecified rather than asserting them to be context-independent; in practice, this gap is usually left unaddressed, and statements default to being read as universally valid. In the
cultural heritage domain, however, statements are rarely neutral facts to
begin with: an attribution, a classification, or a toponym is already the
product of an interpretive act, situated within a particular temporal
period, geographical region, cultural community, political viewpoint,
linguistic tradition, or other epistemic framework. From an epistemological
perspective, this distinction corresponds to the difference between
knowledge that is considered universally valid --- such as the claim that
the capital of Japan is Tokyo --- and knowledge whose validity is contingent
upon a particular observer, community, or worldview. Cultural heritage
objects frequently embody the latter form. While original use cases depended on the former (e.g.\ medical and
scientific domains), cultural heritage embodies the need for the latter
interpretation.

Historical maps, museum catalogues, ethnographic classifications, and
documentary records often reflect the beliefs and conceptual systems of
the societies that produced them. Preserving these artefacts therefore requires preserving not only
factual assertions but also the contexts within which those assertions
were considered meaningful or true --- including, where relevant, our own
modern (or postmodern) frame of interpretation alongside earlier ones.
Otherwise, the insight these artefacts can offer risks being collapsed
into a single, present-day reading, at the cost of the very plurality of
meaning that makes them valuable to cultural heritage. Rather than treating context as an unstructured annotation, we propose that
contextual validity should be represented explicitly through dimensions.

A dimension defines a domain within which the validity of a statement may
vary. Examples include time, space, political perspective, social
identity, religion, language, or scientific paradigm. Each dimension
possesses its own set of permissible values and its own semantics for
reasoning and retrieval. For example, temporal values may be represented
as intervals (OWL-Time), spatial values as geographic
regions (GeoSPARQL), and socio-cultural values as
communities or social groups represented and/or described through taxonomies or ontologies. By making
dimensions explicit, contextual validity becomes a first-class component
of the knowledge model rather than an application-specific interpretation
of metadata.

\subsection{Formal MKG Model}\label{sec:formal-model}

A Multidimensional Knowledge Graph (MKG) extends a conventional knowledge
graph by associating statements with contextual validity constraints. Let
$G=(E,R,T)$ be a knowledge graph consisting of entities ($E$), relations
($R$), and statements ($T$), where each statement is a triple
$t=(s,p,o) \in T$. An MKG associates a statement $t$ with one or more
contexts $C_t = \{c_1,c_2,\ldots,c_n\}$. Each context is defined as a pair
$c=(d,\kappa)$ where $d$ is a dimension and $\kappa$ is a validity
constraint defined over the value space of that dimension. A dimension
$d$ is characterised by a value domain $V_d$. Examples include continuous
data-time values for temporal dimensions or the WGS~84 coordinates system
for global spatial dimensions. Operationally, constraints are represented using a restricted set-theoretic algebra. A constraint restricts which functions 
$\kappa : V_d \rightarrow \{\mathit{true},\mathit{false}\}$ are admissible relative to a declared subset $S \subseteq V_d$, without determining $\kappa$ uniquely (except for Excludes(S)).  
The five core operators are:
\begin{small}
\begin{itemize}
\item $\mathrm{Only}(S)$: The value set contains only members of $S$; nothing outside S is admitted;
\item $\mathrm{Excludes}(S)$: None of the values in the value set belong to $S$;
\item $\mathrm{AnyOf}(S)$: At least one value in the value set belongs to $S$;
\item $\mathrm{AllOf}(S)$: Every value in $S$ is present in the value set; values outside $S$ may also be present;
\item $\mathrm{Exactly}(S)$: The value set coincides exactly with $S$;
\end{itemize}
\end{small}
We illustrate these operators on a scholarly-attribution dimension drawn
from the Shakespeare authorship question: $V_d = \{\text{Stratfordian},
\text{Oxfordian}, \text{Baconian}, \text{Marlovian}, \text{Derbyite}\}$,\\
with $S = \{\text{Oxfordian}, \text{Baconian}, \text{Marlovian},
\text{Derbyite}\}$ denoting the anti-Stratfordian schools, as against the
Stratfordian mainstream view that the plays were written by the
glover's son from Stratford-upon-Avon. ``The author was not William
Shakespeare of Stratford'' with $\mathrm{Exactly}(S)$ holds for exactly
the four anti-Stratfordian schools. ``The author had no more than a
grammar-school education'' with $\mathrm{Excludes}(S)$ is rejected by
all four, without determining whether it holds elsewhere. $\mathrm{Only}(S)$
is illustrated by the claim of a deliberately concealed authorial
identity: endorsed only within $S$, so $\{\text{Oxfordian},
\text{Baconian}\}$ satisfies it but $\{\text{Oxfordian},
\text{Stratfordian}\}$ does not. $\mathrm{AllOf}(S)$ is illustrated by
the claim of a legal and courtly education beyond grammar-school
Stratford, held by all four schools and possibly also by some
heterodox Stratfordians. Finally, $\mathrm{AnyOf}(S)$ is illustrated by
the claim of a cipher naming the true author in the sonnets' dedication,
satisfied as soon as one school in $S$ endorses it --- as the Baconians
famously have.

\subsection{Representation in RDF and SHACL 1.2}\label{sec:rdf-shacl}

The MKG is designed to remain fully compatible with RDF 1.2, which allows metadata to be attached directly to individual statements through quoted triples 
(expressed without being asserted), associated with a MKG context resource that defines the dimension and constraint:
\begin{small}
\begin{verbatim}
<< cskg:NorthCyprus rdf:type cskg:Country >>
    mkg:hasContext _:ctx1 .
_:ctx1
    mkg:dimension cskg:PoliticalPerspective ;
    mkg:constraint _:c1 .
_:c1
    mkg:operator mkg:Exactly ;
    mkg:valueSet ( cskg:Türkiye ) .
\end{verbatim}
\end{small}
In this representation, the quoted triple remains an ordinary RDF
statement, while the contextual information is expressed through RDF
resources that describe the dimension and the associated validity
constraint. Different dimensions may define specialised value types and
reasoning mechanisms. Temporal dimensions may operate over intervals,
spatial dimensions may integrate GeoSPARQL geometries and topological
relations, and social or political dimensions may exploit ontological
hierarchies through OWL-RL reasoning.

The dimension specifies the domain of variation along which the validity
of a statement differs. Beyond the temporal, spatial, and political-perspective dimensions used as
running examples in this paper, we anticipate the MKG accommodating a
broader set of dimension types relevant to cultural heritage metadata,
including opinion and interpretation, narrative framing, bias, legal reuse
status, provenance, uncertainty, and workflow stage --- some of which
(e.g., bias, provenance) are better modelled as properties cutting across
multiple dimensions rather than as dimensions in their own right. We return in Section.~\ref{sec:discussion} to why we do not attempt a full taxonomy of dimensions here, and instead leave their adoption to organic convergence among user communities.

The constraint defines a set
of values for which the statement is valid as a subset of all possible
values for that dimension's value space. Because each context resource is
attached to a quoted triple's reifier, the MKG's dimension and constraint
structures can be validated using SHACL 1.2's new reification constraints\footnote{\url{https://w3c.github.io/data-shapes/shacl12-overview/}}:
\texttt{sh:reifierShape} and \texttt{sh:reificationRequired}. 

These chain a shape on a base triple to a second shape constraining its reifier:
\begin{small}
    
\begin{verbatim}
mkg:ContextualizedStatementShape
    a sh:PropertyShape ;
    sh:reificationRequired true ;
    sh:reifierShape mkg:ContextShape .
mkg:ContextShape
    a sh:NodeShape ;
    sh:property [ sh:path mkg:dimension ; sh:minCount 1 ] ;
    sh:property [ sh:path mkg:constraint ; sh:node mkg:ConstraintShape ; sh:minCount 1 ] .
\end{verbatim}

\end{small}
This guarantees that every statement declared as contextualized carries a
well-formed \texttt{mkg:dimension}/ \texttt{mkg:constraint} pair, giving
the model an enforceable, checkable counterpart rather than a
documentation convention. SHACL 1.2's recently proposed Node Expressions\footnote{\url{https://www.w3.org/TR/shacl12-node-expr/}} further allow context-dependent resolution --- e.g., dynamically
selecting the value of \texttt{cskg:preferredName} appropriate to a
queried \texttt{mkg:dimension} value at query time --- though we note this
specification is at First Public Working Draft stage and not yet
implemented in production SHACL engines. The next section illustrates the
MKG with a concrete cultural heritage case.

\section{Examples}\label{sec:examples}

To demonstrate how MKG-based querying can support context-aware results within the
existing RDF and SPARQL 1.2 specifications, we take an example which is
common in cultural heritage: unclear or contested authorship of
a historical document. The Donation of Constantine was believed by
Western Christendom for centuries to be a decree issued by the Roman
Emperor Constantine granting Pope Sylvester I temporal authority over the
Western Roman Empire. However, in 1440 the scholar Lorenzo Valla used
linguistic and historical analysis to assert that the
document could not have been written at that time and must be a later
forgery (by an unknown cleric, likely in the 8th century).

This case exhibits the kind of contested, time-indexed knowledge
the MKGs are designed to preserve: the document's authorship is not simply
uncertain, but differently certain depending on the consensus of medieval Western
Christendom or modern critical historiography (Valla's 1440 analysis). We can model the original belief across multiple dimensions: it was considered true
between c.750 and 1440, within the political viewpoint of Papal
Supremacy, by Western Christendom and its
validity was supported by the authority of the Catholic Church. For
brevity, we show only two dimensions in the example:
\begin{small}
    
\begin{verbatim}
<< cskg:DonationOfConstantine cskg:hasAuthor cskg:ConstantineI >>
    mkg:hasContext _:ctxMedievalBelief, _:ctxPoliticalBelief, _:ctxSocietalBelief, _:ctxAuthorityBelief .

_:ctxMedievalBelief
    mkg:dimension cskg:TemporalDimension ;
    mkg:constraint _:cMedievalTemporal .

_:cMedievalTemporal
    mkg:operator mkg:Exactly ;
    mkg:valueSet ( cskg:MedievalTimeInterval )

cskg:MedievalTimeInterval
    a time:Interval ;
    time:hasBeginning [ time:inXSDgYear "0750"^^xsd:gYear ] ;
    time:hasEnd       [ time:inXSDgYear "1440"^^xsd:gYear ] .

_:ctxAuthorityBelief
    mkg:dimension cskg:InstitutionalDimension ;
    mkg:constraint _:cMedievalAuthority .

_:cMedievalAuthority
    mkg:operator mkg:Exactly ;
    mkg:valueSet ( cskg:CatholicChurch )
\end{verbatim}

\end{small}
This structure supports context-aware retrieval directly in SPARQL 1.2,
without any extension beyond quoted-triple pattern matching. It may be
noted that spatial and temporal dimensions benefit from spatial
(GeoSPARQL) and temporal (OWL-Time) reasoning. The other dimensions rely
on Set Theory, with or without a formal definition of the permitted
values. The following query retrieves the believed author of the
Donation according to whose authority --- note that this checks only the
context of the institutional authorities that supported the claim
according to the definition of \texttt{InstitutionalDimension} (the value
set of all Organizations). For completeness, one could also ask for the
temporal limits of that belief (if given, so as \texttt{OPTIONAL} in
SPARQL). Note that the set theoretic approach supports establishing the
claim in combination with an ontological representation of the value
space of the dimension (e.g.\ if we ask for the authorship according to
any Christian Church, we should get the claim of the Catholic Church) as
well as without, e.g. if we ask for authorship according to the
Protestant Church, the default result is NIL as we only know the claim of
the Catholic Church - something that also functions without a formal definition of the value space, as basic set logic says that the RDF node for Protestant Church is not in a set which is Only(Catholic Church).
\begin{small}
\begin{verbatim}
SELECT ?author ?authority WHERE {
  << cskg:DonationOfConstantine cskg:hasAuthor ?author >>
     mkg:hasContext ?ctx .
  ?ctx mkg:dimension cskg:InstitutionalDimension ;
       mkg:constraint ?c .
  ?c mkg:operator mkg:Exactly ;
     mkg:valueSet/rdf:rest*/rdf:first ?authority .}
\end{verbatim}
\end{small}
The longer term query model for MKGs could be the definition of custom
functions as SPARQL extensions which are specific to dimension types based on the set-theoretic algebraic operators. As
mentioned before, most dimensions apart from those with clearly known
value spaces that can reuse standardised value types (e.g.\ spatial,
temporal) will benefit from formally defined domain ontologies though
this may lead to new ontological disagreements. A benefit of the Linked
Data / RDF approach is how users only need to define what they need, and
as ontologies for different domains are created, they can be shared
online and re-used. However, the conceptual model of MKGs is also applicable without any ontological commitment to any specific dimensions and values.

\section{Discussion: Benefits and Challenges}\label{sec:discussion}
The MKG model offers three principal benefits over context-independent
knowledge graph representations, as outlined: it preserves
multiple coexisting interpretations without forcing a single authoritative
statement; it enables context-aware querying, allowing applications to
retrieve the version of a statement appropriate to a declared viewpoint;
and it makes contested or evolving knowledge --- as in the example (Section.~\ref{sec:examples}) --- a structured, queryable property of the
graph rather than a fact buried in free-text commentary.

These benefits come with the costs anticipated in \cite{Orlandi2021BenchmarkingRM,pasqual2024evaluation}. Attaching a context to every potentially contested statement increases the size and structural complexity of the graph, with consequences for query performance that we have not yet benchmarked. The model also decides to interpret the logical relation between co-occurring contexts on a statement as conjunctive: e.g. in the Donation of Constantine example (Section.~\ref{sec:examples}) all four contexts hold together --- yet nothing in the RDF encoding distinguishes this from a disjunctive reading over repeated \texttt{mkg:hasContext} triples. RDF itself is agnostic on this point: the conjunctive default is a convention external to RDF's model-theoretic semantics, and any MKG query engine or reasoner must be told to adopt it, rather than inferring it from the graph structure alone. A statement whose validity is genuinely disjunctive across contexts --- valid under context A or under context B, without committing to both holding simultaneously --- cannot rely on this convention, and instead requires the underlying (s,p,o) triple to be asserted as separate occurrences, each carrying its own conjunctive bundle of contexts via a distinct reifier, since RDF 1.2 permits multiple reifiers to be associated with structurally identical triples. This introduces triple duplication in the graph, and stores or pipelines that canonicalize or merge structurally equivalent triples --- a common optimization --- risk collapsing these distinct occurrences back into one, erasing the intended disjunctive partition and silently reintroducing the same AND/OR ambiguity the separate occurrences were meant to resolve.

Defining a dimension's value space may become an ontology engineering task --- the \texttt{cskg:Political- PerspectiveDimension} example in Section.~\ref{sec:formal-model} presupposes an agreed taxonomy of political perspectives, which is
rarely neutral to construct.
Most fundamentally, deciding which dimensions a given
collection should expose, and who is authorized to populate their value spaces, is a
governance question the model does not answer on its own. We take this
openness as deliberate: no central authority can settle, once and for
all, which dimensions a domain needs. Temporal and spatial dimensions
already enjoy a privileged status not because their semantics were
imposed top-down, but because separate communities converged, through
OWL-Time and GeoSPARQL respectively, on representations they
independently found useful. We expect other dimensions relevant to
cultural heritage, such as political perspective or institutional
authority, to stabilise the same way. The MKG's role is to provide the
structural slot into which such conventions can be plugged --- the
dimension/constraint pattern of Section.~\ref{sec:formal-model} --- not
to prescribe their content in advance.

The SHACL 1.2 validation layer introduced in Section.~\ref{sec:rdf-shacl} offers a partial
technical response to this last point. SHACL 1.2 Profiling allows
each community's or institution's accepted set of validity constraints to
be declared as a named \texttt{prof:Profile} of a shared MKG core --- for
instance, a Unionist/Nationalist profile of the Derry/Londonderry
naming (Section \ref{sec:formal-model}), each importing the same base shape from Section \ref{sec:rdf-shacl} but
diverging on which \texttt{mkg:constraint} values are valid.

This makes competing interpretations explicit, citable, and independently
versioned RDF resources rather than implicit application logic, and lets
a consuming application test conformance against any registered profile
without needing to know in advance how many perspectives exist. It does
not, however, resolve who is authorized to register a profile in the
first place, nor does it prevent the proliferation of
contradictory or competing profiles ---
these remain governance questions a validation layer can only make
visible, not settle.

\section{Future Work and Conclusion}\label{sec:conclusion}

The presented MKG model provides a uniform mechanism for representing
contextual validity of statements across arbitrary dimensions of
epistemic validity. By separating statements, dimensions, and constraints,
MKGs enable context-aware querying and reasoning without requiring
modifications to the underlying RDF 1.2 data model, meaning they can be
expressed, stored, queried and used in any tool/system which is RDF 1.2
conform. Building on the governance limitations identified in Section.~\ref{sec:discussion}, we
also propose to design data profiles attached to MKG dimension annotations as
machine checkable expressions of FAIR\cite{wilkinson_fair_2016} and CARE\cite{Carroll-2020} commitments at
statement level. 
Community-sensitive statements may carry authorization
metadata analogous to Local Contexts' Traditional Knowledge Labels\cite{localcontexts_tklabels}
before being resolved outside their originating community's context.
Realizing this will require genuine consultation with the communities
such profiles claim to represent, which a schema alone cannot substitute
for. INFINITY will publish the final MKG concept as a public Note for
discussion by the RDF/Linked Data community and push for adoption within
the context of the use in the ECCCH and DS4CH initiatives, finally
enabling cultural heritage institutions a means to preserve the
contextual interpretations of annotated objects through their metadata.

\section*{Acknowledgments}
This work has received funding from the European Union's Horizon Europe
research and innovation programme under grant agreement N.~101233051
INFINITY project (\url{https://infinity-eccch.eu/}).

\bibliography{sample-ceur}

\end{document}